LOGO

# Synthetic IMU Datasets and Protocols Can Simplify Fall Detection Experiments and Optimize Sensor Configuration

Jie Tang, Bin He, *Member, IEEE*, Junkai Xu, Tian Tan, Zhipeng Wang, Yanmin Zhou, Shuo Jiang*, *Member, IEEE*

**Abstract**—Falls represent a significant cause of injury among the elderly population. Extensive research has been devoted to the utilization of wearable IMU sensors in conjunction with machine learning techniques for fall detection. To address the challenge of acquiring costly training data, this paper presents a novel method that generates a substantial volume of synthetic IMU data with minimal real fall experiments. First, unmarked 3D motion capture technology is employed to reconstruct human movements. Subsequently, utilizing the biomechanical simulation platform Opensim and forward kinematic methods, an ample amount of training data from various body segments can be custom generated. An LSTM model is trained, achieving testing accuracies of 91.99% and 86.62% on two distinct datasets of actual fall-related IMU data, demonstrated the comparable performance of models trained using genuine IMU data. Building upon the simulation framework, this paper further optimized the single IMU attachment position and multiple IMU combinations on fall detection. The proposed method simplifies fall detection data acquisition experiments, provides novel venue for generating low cost synthetic data in scenario where acquiring data for machine learning is challenging and paves the way for customizing machine learning configurations.

## I. INTRODUCTION

FALLING is one of the main causes of injuries among the elderly, exerting profound impacts on their health and quality of life. It is estimated that approximately 684,000 cases of fatal fall injuries occur annually, positioning them as the second leading cause of unintentional injury-related deaths, following closely behind road traffic injuries.

Notably, within all fall-related incidents, individuals aged 60 and above exhibit the highest mortality rates. Elderly individuals face the greatest risk of death or severe injury due to falls, with this risk escalating as age advances. For instance, in the United States, 20% to 30% of elderly individuals involved in fall accidents endure moderate to severe injuries, encompassing abrasions, hip fractures, and cranial trauma, among others [1]. Consequently, the pursuit of fall detection research and its practical implementation hold crucial socio-medical significance, offering the potential to furnish comprehensive care and assurance for the well-being of the elderly.

A diverse array of existing methodologies exists within the realm of fall detection. From the sensing perspective, these can be broadly classified into two categories: camera-based and wearable Inertial Measurement Unit (IMU) based approaches [2]. In contrast to camera-based fall detection systems, wearable IMU-based systems are characterized by their ease of deployment and immunity to occlusion challenges. Early implementations of fall detection systems utilizing IMU sensors employed simplistic threshold-based detection algorithms, whereas the feasibility of machine learning techniques in enhancing the recognition of fall movements has been demonstrated [3][4]. Wearable devices, particularly those incorporating IMUs, in combination with machine learning methodologies, play a pivotal role in fall detection. Queralta et al. established a fall detection system encompassing IMU sensor nodes, edge gateways, cloud services, and end-user applications [5]. Liu et al. fused IMU sensor networks with Doppler radar systems, resulting in a significant reduction in false positive rates within fall detection systems [6]. He et al. integrated IMU sensors into a vest tailored for the elderly,

Manuscript received XX, 2023; This work was supported by the National Natural Science Foundation of China under Grant 61825303, 52105033, and 62088101; in part by Shanghai Municipal Science and Technology Shanghai Sailing Program under Grant 21YF1450400 and Major Project under Grant 2021SHZDZX0100; in part by Chenguang Program by Shanghai Municipal Education Commission under Grant 21CGA23. (*Corresponding authors: Shuo Jiang.)

Jie Tang, Bin He, Zhipeng Wang, Yanmin Zhou and Shuo Jiang are with the College of Electronics and Information Engineering, Tongji University, Shanghai 201804, China, and also with Frontiers Science Center for Intelligent Autonomous Systems, Shanghai 200120, China (e-mail: tangjie1953479@tongji.edu.cn, hebin@tongji.edu.cn, wangzhipeng@tongji.edu.cn, yanmin.zhou@tongji.edu.cn, jiangshuo@tongji.edu.cn).

Junkai Xu is with the State Key Laboratory of Mechanical System and Vibration, School of Mechanical Engineering, Shanghai Jiao Tong University, Shanghai 200240, China (e-mail: xujunkai@imumaster.com).

Tian Tan is with the Department of Radiology, Stanford University, Stanford, CA, USA. (e-mail: alanttan@stanford.edu).

transmitting data via Bluetooth to a mobile smartphone running a fall detection program based on the k-Nearest Neighbors (KNN) algorithm, thereby enabling automatic alerts upon fall incidents [7]. Xu et al. introduced a fall detection algorithm founded on a CNN-LSTM (Convolutional Neural Network-Long Short-Term Memory) composite network. This approach leverages neural networks to automatically extract features and conduct classification, obviating intricate preprocessing procedures [8].

The utilization of wearable IMUs in conjunction with machine learning for fall detection necessitates a substantial quantity of IMU data encompassing both fall and non-fall activities. However, the cost of acquiring fall detection data is high. Conventional data collection approaches typically involve tracking the IMU data of elderly individuals or recruiting younger participants to perform motion experiments for data collection. Lisowska et al. collected data from 20 volunteers wearing Silmee devices during simulated falls and daily activities [9]. Some researchers have publicly released experimental datasets, such as the MobiFall dataset [10] and the SisFall dataset [11]. Bagala et al. established the FARSEEING database, noting that a comprehensive shared real-world fall database could enhance the understanding of fall processes [12].

Gathering IMU data through real-world experiments necessitates addressing two vital pre-experiment considerations: experimental protocol design and IMU sensor placement. Once the experimental protocol is established, post-experiment parameter adjustments become challenging, often requiring costly reiterations of the experiments. Despite comprehensive experimental protocol designs being proposed to contrast fall and non-fall movements [13], variations in action designs persist across experiments and datasets. This diversity may stem from considerations of experimental complexity or differing research foci. Özdemir et al. included 20 fall movements and 16 non-fall movements in their experiments [14], whereas Alfred Wertner's research examined only 4 fall and 10 non-fall movements [15]. Wang et al. highlights that the datasets utilized in most papers often fail to encompass all types of falls found in real-world scenarios [16]. Variations in IMU quantities and placement positions are also evidently different across different datasets. Tong et al. propose that the wrist, hip, and thigh are not suitable positions for accelerometer-based devices due to their high frequency and complexity of motion. Instead, they suggest that regions below the neck and above the waist, specifically the upper torso, are more appropriate for distinguishing fall-related characteristics using acceleration data [17]. Nukala et al. designed an artificial neural network for fall detection using IMU data from the waist and back, achieving a sensitivity of 97.0% and specificity of 97.2% [18]. In the experiments conducted by Ojetola et al., two SHIMMER sensors were attached to the participant's chest and thigh [19]. Yan et al. proposed a fall detection method based on multi-directional sensor skeletons using ST-GCN and employed the Up-Fall dataset [20], which places IMUs on the neck, waist, thigh, wrist, and ankle for model training and testing [21]. In summary, the requirement for participants to wear IMU devices during motion experiments for data collection is associated with both high costs and limited flexibility. Furthermore, the integration of these datasets presents challenges due to variations in the experimental designs across different studies, potentially resulting in a diminution of data set quality.

To address data limitations, researchers have proposed various approaches: (1) Data Augmentation Techniques: Researchers have employed techniques like rotation to augment limited fall-related IMU datasets. Theodoridis et al. augmented training data by rotating the original data around random angles within the range of [-10, 10] degrees, enhancing the model's robustness to IMU rotations [22]. Yhdego et al. inverted the sign of sensor data across all three axes to simulate sensor inversion [23]. (2) Transfer Learning Methods: Butt et al. utilized transfer learning by retraining a VGG-16 convolutional neural network (CNN), initially trained on a large-scale image dataset, to learn features relevant to human actions for fall detection tasks [24]. Yhdego et al. repurposed the AlexNet architecture of deep convolutional neural networks, modifying and retraining the last three layers of the AlexNet architecture. They employed wavelet-based time-frequency analysis to represent sensor data through scaleograms, which were then fed into the AlexNet framework [23]. (3) Biomechanical Simulation Tools for Synthetic Data: Mastorakis et al. utilized Opensim simulation software to generate synthetic data. They extracted a person's height and body orientation before a fall from depth camera-captured videos. These pieces of information are utilized to tailor the height of skeletal models and the falling motion, thereby generating vertical acceleration at the center of mass [25].

Among these approaches, the biomechanical simulation-based method has gained increasing attention from researchers and has seen successful applications in other research domains. Recinos et al. utilized an integrated framework of Opensim and Unity to estimate ground reaction forces in a simulation environment using IMU data collected from real motion experiments [26]. Jiang et al. used fatigue detection motion experiment data from a limited number of participants to drive model movements in Opensim. Simulated individuals with different physiological attributes executed movements, thereby enhancing the IMU dataset. Their research demonstrated that this model-based data augmentation approach outperformed traditional methods such as rotation, random noise, amplitude warping, time warping, scaling, and proposed methods [27]. Uhlenberg et al. introduced a joint simulation framework consisting of a biomechanical human model and wearable inertial sensor models. A total of 960 inertial sensors were virtually connected to the lower limbs and shoe model of the human body in Opensim. They simulated gait patterns of hemiparetic patients using motion capture data, generating synthetic accelerations and angular velocities based on the inertial sensor model, enabling comprehensive analysis of gait events [28]. The aforementioned studies have substantiated the feasibility of employing simulation techniques to synthesize IMU data. This approach holds potential significance for addressing fall detection challenges in scenarios characterized by substantial data paucity.

As per our investigation, existing real IMU datasets in the field of fall detection suffer from issues such as limited data volume and inconsistent experimental paradigms. In contrast, an abundance of video datasets capturing falls is available, offering the potential to replicate more realistic fall scenarios. Additionally, the feasibility of utilizing biomechanical simulations to generate synthetic IMU data has been established and met with successful application, while its potential remains underutilized in the domain of fall detection. Considering the above factors collectively, the integration of vision-based 3D motion reconstruction with Opensim biomechanical simulation enables cross-modal data generation (from video to IMU). This methodology holds promise within the domain of fall detection, as it effectively addresses the challenge of scarce fall-specific data availability. However, research on this approach remains scarce.

To address the challenges of acquiring costly fall training data, optimizing sensor positions and combinations, this paper proposes a methodology for fall detection that leverages biomechanical simulation software, such as Opensim, to generate synthetic IMU data for training machine learning models. The resulting models exhibit favorable performance on real fall experiment IMU datasets. The contributions of this paper are summarized as follows:

1. Cost-effective Data Generation: In response to the high cost of fall-related data, this paper employs Opensim to generate synthetic IMU data of human movements reconstructed from video. The dataset encompasses various fall and non-fall movements. By manipulating model dimensions and IMU placement poses, the methodology enhances synthetic data through data augmentation, yielding a robust synthetic dataset.

2. Customized IMU positions: Through simulation, this study customizes the generation of synthetic IMU data from distinct body regions, investigating the impact of IMU placement on fall detection. The effectiveness of the proposed methodology is validated from both signal and model perspectives.

3. Optimized the Combination of Multiple IMUs: Based on simulation, the combination of synthetic IMU data from multiple body regions has been optimized to provide a comprehensive representation of falling actions.

## II. METHODS

### A. Overview

The system is broadly divided into three main components: driving Opensim skeletal models, generating synthetic IMU data, and training/testing machine learning models (Fig. 1). OpenSim is a sophisticated biomechanical modeling and simulation software platform that enables the creation and analysis of musculoskeletal models to study human movement and biomechanics [29]. Initially, to drive Opensim skeletal models, multiple cameras capture human fall actions, followed by OpenPose detecting key body points in the videos. The three-dimensional coordinates of these key points are calculated using the triangulation algorithm. Subsequently, human movements are reconstructed via Opensim's marker-based inverse kinematics. Simulated IMUs can be placed freely on the skeletal model, and thus the synthetic IMU raw data can be extracted during model movements. Following this, multiple machine learning models, such as LSTM networks, are trained using the synthetic IMU data and tested on both publicly available IMU datasets and experimentally collected IMU datasets to demonstrated the feasibility of the proposed methods.

### B. 2D Pose Estimation

Traditional methods for reconstructing human motion based on optical markers rely on expensive optical motion capture equipment. To achieve markerless motion reconstruction without specialized equipment, this study employs multiple cameras to record human motion videos. Human body pose estimation is performed on the recorded videos from each camera, yielding two-dimensional pixel coordinates of human keypoints. OpenPose, a widely used 2D human pose estimation algorithm, accomplishes pose detection by analyzing human keypoints within the images [30]. It employs a multi-stage deep learning architecture, encompassing a convolutional neural network (CNN) stage for body part detection and a pose association stage for linking body parts. In the initial stage, CNN is employed to detect body parts in the images, such as the head, hands, and feet. The network employs convolution, pooling, and fully connected layers to extract features, generating confidence maps and positional information associated with each keypoint. In the subsequent stage, keypoint confidence maps generated in the preceding stage are utilized to associate body parts. This stage involves generating part affinity fields to describe the associational information connecting various body parts. By detecting associations between connected body parts, the algorithm discerns the structure of the human pose. Upon obtaining preliminary keypoint and association information, refinement and filtering of detected poses ensue. Precise estimation of body part positions is achieved by leveraging associational information based on connected body parts, followed by thresholding and non-maximum suppression to eliminate redundant keypoints. The ultimate output comprises coordinates and confidences of detected human keypoints, alongside associated connection information, which collectively represent the pose of the human body in the image.

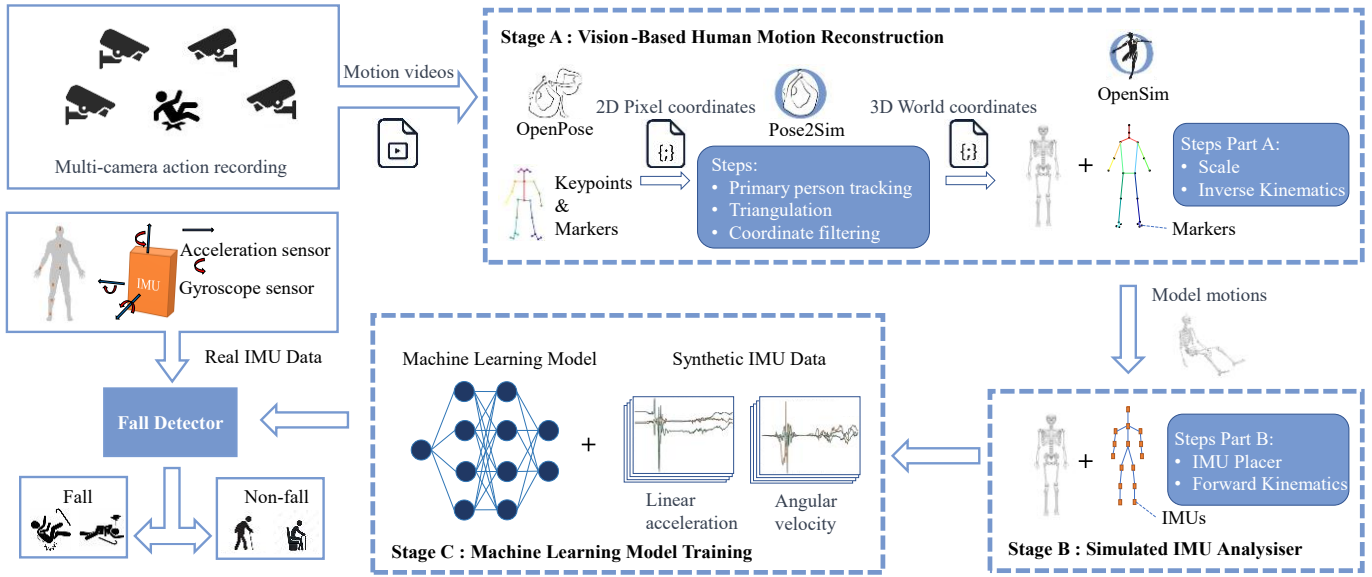

Fig. 1. Experimental framework of the proposed system.

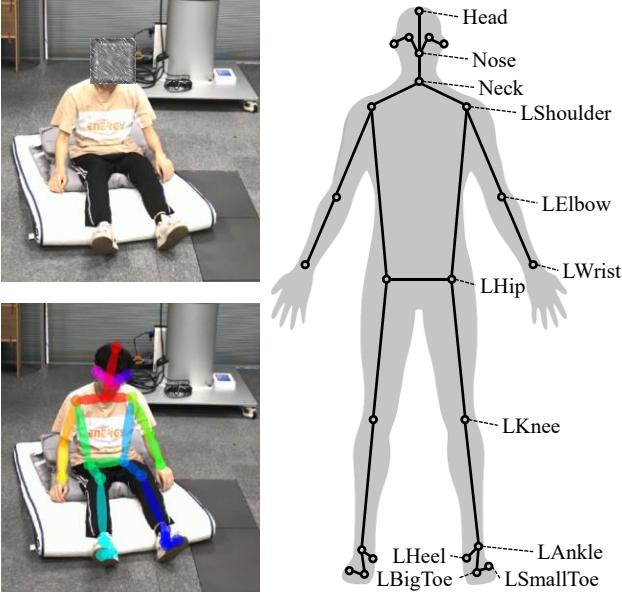

**Fig. 2.** 2D human pose estimation in video. The Body_25B model, employed for human pose detection, encompasses a total of 25 distinct keypoints of the human body.

### C. 3D Coordinate Calculation

The marker-based inverse kinematics solving algorithm of Opensim requires the three-dimensional world coordinates of all marker points. This necessitates camera calibration to determine the camera's pose relative to the ground. Subsequently, the pixel coordinates of key points in camera videos are converted to world coordinates. This paper employed the classical chessboard calibration method [31]. Pose recognition is initially performed on the human motion videos recorded by each camera. Leveraging the multi-camera intrinsic and extrinsic parameters obtained during camera calibration, a triangulation method converts the 2D coordinates of human key points from different camera views captured at the same moment in a video frame into 3D coordinates in the world coordinate system. Finally, a Gaussian filter is applied to smooth the 3D coordinates of all key points, aiming to refine the motion obtained through the inverse kinematics calculation step.

Direct linear transformation is a classic triangulation algorithm, and incorporating weighted confidence of each keypoint provides an effective enhancement to this method [32]. Let $Q = (X, Y, Z, 1)^T$ be the three-dimensional homogeneous coordinates of a keypoint in the world coordinate system, $q = (u, v, 1)^T$ represents its two-dimensional homogeneous coordinates in the pixel coordinate system, $P = (P_1, P_2, P_3)^T$ is the camera projection matrix obtained from the camera calibration process, $\lambda$ is the scale factor, and c is the confidence of the keypoints. According to camera geometry theory, the following equation can be deduced:

$$PQ = \lambda q \tag{1}$$

Multiplying both sides of the equation by $q$ , we get:

$$q \times PQ = 0 \tag{2}$$

Utilizing the confidence of keypoints for weighting:

$$c \times (P_1 - uP_3)Q = 0$$
$$c \times (P_2 - vP_3)Q = 0 \tag{3}$$

For N cameras, there are 2N equations, which can be represented as:

$$AQ = 0 \tag{4}$$

Using the Singular Value Decomposition (SVD) of matrix A to solve for the least squares solution of matrix Q:

$$A = USV^T$$
$$Q = (V_{14}/V_{44}, V_{24}/V_{44}, V_{34}/V_{44}, 1) \tag{5}$$

### D. Inverse Kinematics Calculation

This step aims to drive the model to perform human actions of interest. Specifically, it involves deducing joint movements that can generate a given motion trajectory. This problem is formulated as an optimization task, where joint angles are adjusted to minimize the discrepancy between simulated and measured trajectories, as depicted in (6). Due to the highly nonlinear nature of this optimization problem, numerical optimization methods are employed for its solution. For solving this least squares problem, a general quadratic programming solver is utilized with a convergence criterion set at 0.0001 and a maximum iteration limit of 1000 iterations [33].

Describing the motion of Opensim skeletal models involves employing joint angles and translational transformations, denoted as generalized coordinates $q$ . Here, $x_i^{\exp}$ represents the three-dimensional coordinates of the marker points obtained through experiments, while $x_i(q)$ signifies the three-dimensional coordinates of simulated marker points at the given $q$; $q_j^{exp}$ denotes the generalized coordinate values obtained from experiments, and $w_i$ represents the weights assigned to the marker points. Therefore, the problem addressed by the inverse kinematics computation is a weighted least squares problem:

$$\min_q \left[ \sum_{i \in markers} w_i \|x_i^{exp} - x_i(q)\|^2 + \sum_j \omega_j (q_j^{exp} - q_j)^2 \right] \quad (6)$$

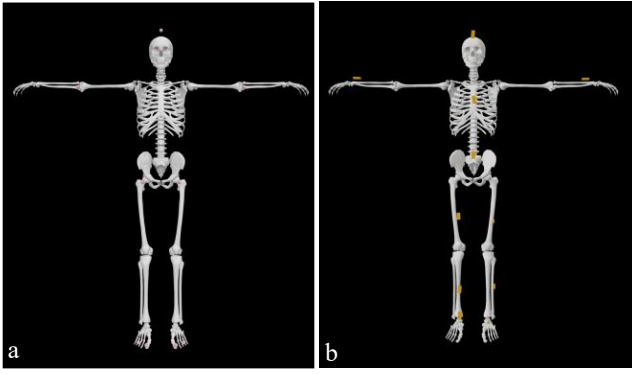

Fig. 3. a. The Body_25B skeletal model of the human body. The pink spheres at the joints in the figure serve as both key points for human pose detection and virtual markers for inverse kinematics calculations. b. Simulated IMU placement. The orange cuboids rigidly attached to the skeleton represent the simulated IMUs.

### E. Forward Kinematic Analysis

Forward kinematic analysis is a method used to determine the motion of specific segments within a biomechanical model given certain joint angles. In this analysis, the focus is on calculating the positions, velocities, and accelerations of various segments of the model during motion, starting from an initial state. Initially, the joint angles and translational transformations at the initial moment are provided, determining the model's initial state. Subsequently, based on the joint angles and the model's geometric structure, the position, angular velocity, and acceleration of the segments of interest are solved using the Newton-Euler equations. Over multiple time steps, the calculations are iterated from the initial pose, ultimately providing the kinematic information of the segments for the given joint angles throughout the entire motion.

Taking a simple two-link system as an example, the angular velocity of the next link is the sum of the angular velocity of the previous link and the angular velocity introduced by the joint's rotation. Let $\theta$ be the angle of rotation joint, $n$ be the rotation axis, and $L$ be the position vector between the links. The angular velocity and acceleration transmission formula for the link system can be expressed as:

$$w_{i+1} = w_i + \dot{\theta}_{i+1} \cdot n_{i+1} \quad (7)$$
$$\dot{v}_{i+1} = \dot{v}_i + \dot{w}_i \times L_{i+1} + w_i \times (w_i \times L_{i+1})$$

To calculate acceleration, angular velocity, and other relevant information of various body segments during human motion, Opensim employs a more refined forward kinematic analysis algorithm. By customizing the placement of simulated

IMUs, it is possible to obtain three-axis acceleration and angular velocity data for that position during the motion of the skeletal model.

### F. Fall Detection Model Training

After obtaining the required synthetic IMU data, the generated synthetic dataset is used for model training. Considering the sequential nature of IMU outputs such as angular velocity and acceleration data, models like Recurrent Neural Network (RNN) are well-suited, particularly recurrent neural networks (RNNs) like LSTM networks. Additionally, this study explored other models such as K-Nearest Neighbors (KNN), Support Vector Machines (SVM), Convolutional Neural Networks (CNN), among others. Among all these models, LSTM demonstrated the highest accuracy and robustness.

LSTM, a variant of RNN, is designed for modeling sequential data and addresses the vanishing gradient problem in traditional RNN. The fundamental structure of LSTM includes the cell state, input gate, forget gate, and output gate [34][35].

Firstly, the cell state is at the core of the LSTM network and is responsible for carrying long-term information. It can be thought of as a channel that runs through the entire network. At each time step, LSTM updates the cell state using the input data and the output from the previous time step through the input gate and the forget gate.

Secondly, The input gate decides which information to add to the cell state. It first computes a candidate value based on the input data and the output from the previous time step. Then, it uses logistic regression to determine which parts of the cell state should be updated. The formula for the input gate is as follows:

$$i_t = \sigma(W_{ix}x_t + W_{ih}h_{t-1} + b_i) \quad (8)$$

Following that, the forget gate decides which information in the cell state to forget. Similar to the input gate, the forget gate employs logistic regression to determine the degree of forgetting for each part. The formula for the forget gate is as follows:

$$f_t = \sigma(W_{fx}x_t + W_{fh}h_{t-1} + b_f) \quad (9)$$

The cell state is updated by multiplying the candidate value from the input gate and the forgotten parts of the cell state, and then adding new information. The formula for updating the cell state is as follows:

$$C_t = f_t \odot C_{t-1} + i_t \odot \tanh(W_{cx}x_t + W_{ch}h_{t-1} + b_c) \quad (10)$$

Lastly, the output gate decides which information will be used for the current time step's output. Like the input gate and forget gate, the output gate employs logistic regression to determine which parts of the output to consider. The formula for the output gate is as follows:

$$o_t = \sigma(W_{ox}x_t + W_{oh}h_{t-1} + b_o) \quad (11)$$

The output state is the final output of the LSTM and is a transformation of the cell state based on the output gate's information and the cell state's transformation. The formula for calculating the output state is as follows:

$$h_t = o_t \odot \tanh(C_t) \quad (12)$$

Through these gating mechanisms and state updates, LSTM

effectively tackles the long-term dependency problem, allowing the network to capture information in sequences better.

In this paper, a Long Short-Term Memory (LSTM) network composed of a Batch Normalization (BN) layer, two LSTM layers, and two fully connected layers is primarily used. The implementation is practically performed using TensorFlow. The training process involves using a cross-entropy loss function and the Adam optimizer with a learning rate of 0.0001. A batch size of 16 and 50 training epochs are employed. The chosen feature is the triaxial acceleration and triaxial angular velocity from IMUs. Each action is represented as a tensor of size 240×N×6 (where N represents the number of IMUs). There are two categories, and one-hot encoding is utilized, where the fall action label is encoded as "01" and the non-fall action label as "10".

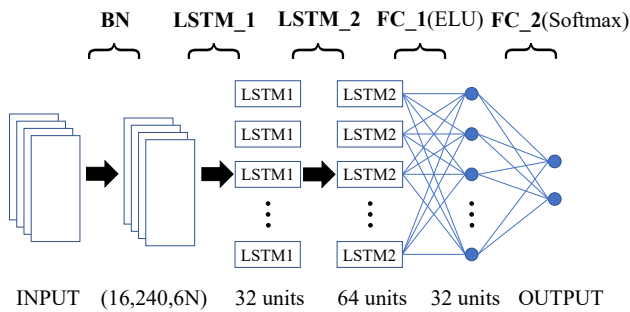

Fig. 4. Structure of the LSTM model.

## III. EXPERIMENTS

### A. Experimental Setups

The experiments were conducted at the Kaiwu Laboratory [36], Tongji University, Zhangjiang AI Island. Kaiwu lab focuses on human-machine interaction datasets and embodied learning algorithms for intelligent robots. An industrial camera system was utilized, with four cameras positioned in the frontal direction. The Luster Capture software was used to synchronize video recording and decompression. The resulting videos had a resolution of 2048×1536, a frame rate of 30 frames per second, and a duration of at least 8 seconds. Seven Xsens MTw2 IMU sensors were affixed to the participants during the experiment, synchronously collecting real IMU data.

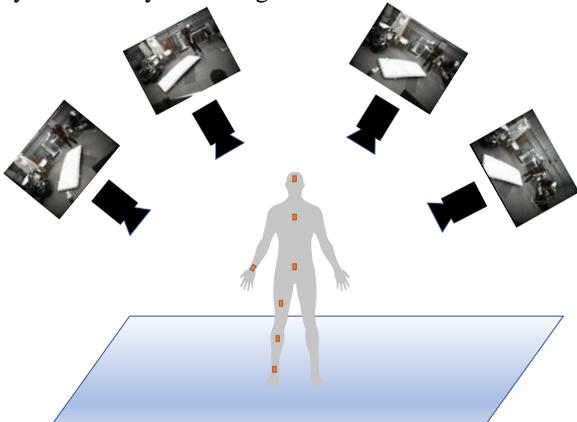

Fig. 5. Experimental setup, using four cameras to capture motion from different angles while wearing 7 IMU sensors.

### B. Experimental Protocol Design

Sixteen fall actions and sixteen non-fall actions were designed (Table 1). Each action was performed three times, with action videos recorded and real IMU data collected. The experimental protocol design drew inspiration from existing research [13][21][37], encompassing both common fall actions and daily activities, while also incorporating some potentially confusing movements.

The present study has been granted ethical approval with the approval number E2021101I. All human participants involved in this study provided informed consent, and their privacy and rights were safeguarded.

TABLE I
ACTION DESIGN (16 CATEGORIES EACH FOR FALLS AND NON-FALLS)

| Non-fall | Fall |
|---|---|
| from standing to lying on the bed | fall forward with arm protection |
| from lying down to sitting | kneel down |
| from standing to sitting on the sofa | kneel down and then lying down |
| from standing to sitting on the chair | falling forward to the right |
| squatting horse stance | falling forward to the left |
| walking | fall forwards and quick recover |
| running | fall forwards and slow recover |
| walking backwards | fall backwards and sit on the floor |
| bending forward at a 90-degree angle | fall backwards and lie on the floor |
| bending over to pick up an object from the ground | fall backwards and make contact with the floor on the right side |
| walking, stumbling, and recovering balance | fall backwards and make contact with the floor on the left side |
| walking with a limp | fall to the right and then lie down |
| from squatting to standing | fall to the right and then recover |
| bending over while walking | fall to the left and then lie down |
| rolling over on the bed | fall to the left side and then recover |
| from sitting to standing, followed by walking forward | roll from lying down on the bed to the floor |

### C. Experimental Parameter Configuration

The primary parameter configuration for Opensim model motion reconstruction based on camera video is as follows. The parameter values are set based on the recommended values from Pose2Sim [32] and further adjusted according to the actual performance of the program during execution.

TABLE II
PARAMETER CONFIGURATION DETAILS

| Parameter Name | Value |
|---|---|
| frame_range | [0,240] |
| frame_rate | 30 |
| pose_model | BODY_25B |
| tracked_keypoint | Head |
| error_threshold_tracking | 20 |
| error_threshold_triangulation | 20 |
| likelihood_threshold | 0.3 |
| interpolation | cubic |
| interp_if_gap_smaller_than | 20 |
| filter_type | gaussian |
| sigma_kernel | 3 |

Based on human motion videos recorded by multiple cameras, filtered human keypoint coordinates were obtained. Using these marker points, inverse kinematics calculations were performed in Opensim to derive model movements. Subsequently, simulated IMUs were placed on different body parts of the model, and forward kinematic analysis was conducted to generate acceleration and angular velocity data for the simulated IMUs.

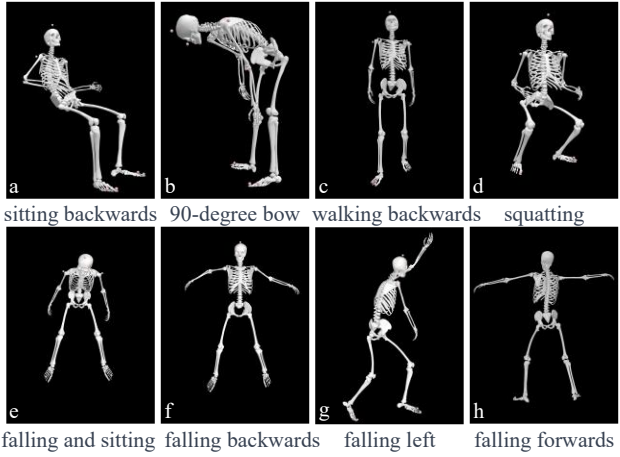

| a | b | c | d |
|---|---|---|---|
| sitting backwards | 90-degree bow | walking backwards | squatting |

| e | f | g | h |
|---|---|---|---|
| falling and sitting | falling backwards | falling left | falling forwards |

Fig. 6. Examples of action reconstruction. In the figure, panels (a), (b), (c), and (d) depict frontal views of four distinct non-falling actions, while panels (e), (f), (g), and (h) illustrate top views of four different falling actions.

### D. Model Training and Testing

Based on the experimentally reconstructed 32 actions, a set of simulation configurations was capable of generating 32 IMU data samples. Data augmentation was performed by scaling the simulation model and adding a small random amount to the pose of each simulated IMU. In total, 32×6 IMU data samples were obtained, which constituted the training dataset.

Two test sets were used in this study, both comprising real fall experiment IMU datasets.

Experimental Dataset A: Seven Xsens MTw2 IMU sensors were worn by participants to collect IMU data (three-axis acceleration and angular velocity, or other data such as Euler angles) during action execution. The dataset involved six participants performing 16 fall actions and 16 non-fall actions, each repeated three times. A single experiment generated 96 IMU data samples, resulting in a total of 513 valid samples. IMUs were attached to the head, chest, waist, front of the right thigh, front of the right shank, front of the right ankle, and right wrist.

Public Dataset B: An authentic dataset of real fall experiments collected by Özdemir et al. was used as a test set [14]. This dataset included 17 participants, 20 non-fall actions, and 16 fall actions, each repeated 5 or 6 times, totaling 3313 fall experiment data samples. Each sample comprised acceleration and angular velocity data from six Xsens MTw wearable IMU sensors, placed on the neck, chest, waist, right thigh, right shank, and right wrist.

### E. Experimental Evaluation Metrics

To verify the feasibility of training fall detection models using synthetic IMU data, a comparison was made between synthetic and real IMU data from both signal and model perspectives.

From the signal perspective, a comparison was conducted between synthetic IMU data and filtered real IMU data collected from the same position during the same action experiment.

From the model perspective, identical models were trained using both synthetic and real data. The performance of these models was tested on the same datasets, and a comparison was made using metrics such as accuracy, precision, recall, and F1-score.

## IV. RESULTS

### A. Synthetic IMU Data

The figure below presents synthetic data generated by Opensim based on video reconstruction of actions. Specifically, a comparison is made between the synthetic IMU data from the waist during a fall action and the corresponding real IMU data. Similarly, a comparison is made between synthetic IMU data from the shank during a non-fall action and the corresponding real IMU data. Overall, the generated simulation data exhibits comparability with the real data at the signal level.

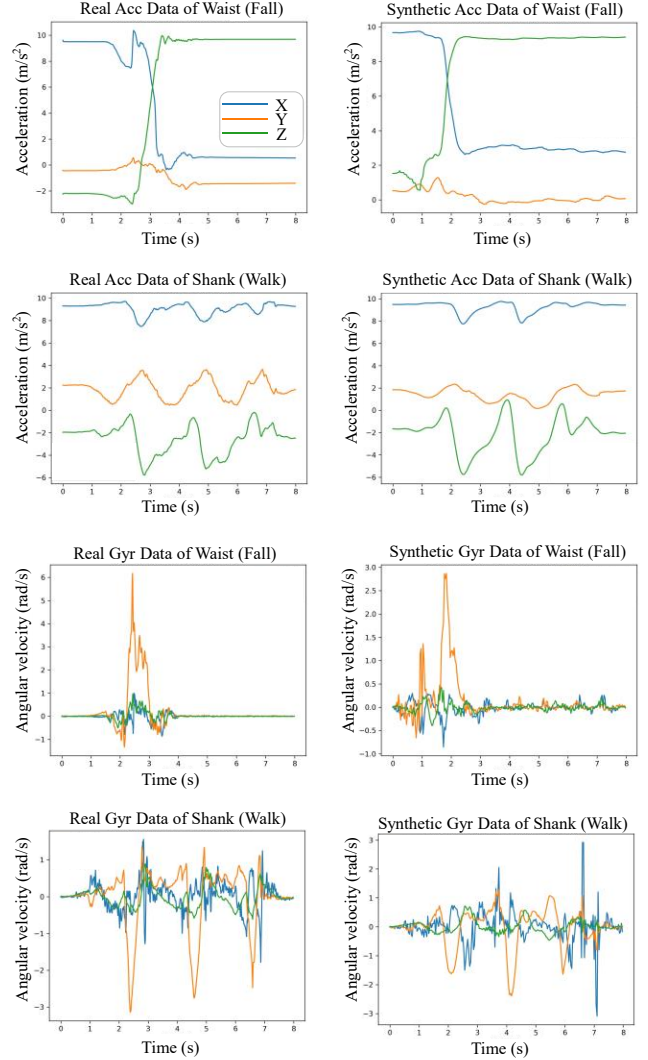

Fig. 7. Comparison of synthetic IMU acceleration data and real data. The first row presents the waist IMU acceleration data comparison for the falling action (falling backward onto the floor), while the second row shows the lower leg IMU acceleration data comparison for the non-falling action (walking backwards). The first column displays real IMU data (filtered to remove high-frequency noise signals), and the second column presents synthetic IMU data.

To facilitate a more comprehensive comparison between synthetic and real acceleration data, Gaussian filtering is applied to the original acceleration data, effectively eliminating high-frequency noise. The resulting curve aligns with the simulated trend. As for the original angular velocity data, no filtering is applied, and while it exhibits a certain degree of resemblance to the synthetic angular velocity data, this similarity is less pronounced compared to the acceleration data. A more in-depth analysis will be conducted in the subsequent discussion section

### B. Model Test Results

The performance of the model trained using synthetic IMU data from the shank on the two datasets is as follows. Additionally, for comparison with models trained on real data, a model was trained using data from one participant on both datasets, and the remaining participants' data were used for testing.

From the table, it is evident that the LSTM model trained using synthetic data achieves an accuracy of 91.99% on Dataset A, which is 4.49% higher than the accuracy of the model trained with real data used as comparison. Additionally, the synthetic data-trained model outperforms the real data-trained model in the other three performance metrics. However, on Dataset B, which encompasses a more extensive set of samples, the accuracy of the model trained with synthetic data drops to 86.62%, resulting in a 0.53% decrease compared to the real data-trained model.

TABLE III
PERFORMANCE OF LSTM MODEL TRAINED WITH LOWER LEG IMU DATA ON DATASET A

|  | accuracy | precision | recall | F1 |
|---|---|---|---|---|
| simulated | **91.99%** | 88.66% | 98.01% | 93.10% |
| real | **87.50%** | 86.22% | 92.80% | 89.39% |

TABLE IV
PERFORMANCE OF LSTM MODEL TRAINED WITH LOWER LEG IMU DATA ON DATASET B

|  | accuracy | precision | recall | F1 |
|---|---|---|---|---|
| simulated | **86.62%** | 89.39% | 86.02% | 87.64% |
| real | **87.15%** | 89.57% | 87.40% | 88.16% |

### C. Misclassification Rates for Different Actions

Using synthetic three-axis acceleration data from the shank as features, the LSTM model was tested on the dataset B. The misclassification rates for different actions are calculated and presented in the table below. For comparison, the results of the model trained on real data are also provided. The results indicate that the four actions with the highest misclassification rates on the dataset B are transitioning from standing to lying on a bed, transitioning from lying down to sitting, falling to the ground on the left side and recovering quickly, and falling and quickly recovering. Similarly, during testing on Dataset A, the results indicate that the three non-fall actions—transitioning from standing to lying on the bed, transitioning from supine to sitting, and rolling over in bed—are prone to being misidentified as fall actions. For other actions, the model demonstrates accurate classification in the majority of cases.

TABLE V
MISCLASSIFICATION RATES ON DIFFERENT ACTIONS

| action | simulated | real |
|---|---|---|
| walking forwards | 0 | 0 |
| walking backwards | 0 | 0 |
| jogging, running | 0 | 0 |
| squatting, then standing up | 7.95% | 0 |
| bending about 90 degrees | 0 | 0 |
| bending to pick up an object on the floor | 0 | 0 |
| walking with a limp | 0 | 0 |
| stumbling with recovery | 0 | 0 |
| bending while walking | 0 | 0 |
| coughing or sneezing | 0 | 0 |
| sitting onto a chair (hard surface) | 0 | 0 |
| sitting onto a sofa (soft surface) | 0 | 0 |
| Squatting horse stance | 21.59% | 4.88% |
| from vertical to sitting onto a bed | 0 | 0 |
| **from vertical to lying on the bed** | **100%** | **100%** |
| **from lying to sitting** | **100%** | **100%** |
| falling forward to the floor | 0 | 0 |
| falling to the floor with arm protection | 0 | 0 |
| falling down on the knees | 0 | 0 |
| falling down on the knees and then lying | 0 | 0 |
| **falling on the floor and quick recovery** | **41.86%** | **41.77%** |
| falling on the floor and slow recovery | 9.09% | 0 |
| falling forward to the right | 0 | 0 |
| falling forward to the left | 0 | 0 |
| falling backward, ending sitting | 3.75% | 11.39% |
| falling backward, ending lying | 3.45% | 0 |
| falling backward to the right | 1.12% | 1.14% |
| falling backward to the left | 1.95% | 0 |
| falling to the right, ending lying | 0 | 0 |
| falling to the right and recovery | 3.45% | 23.08% |
| falling to the left, ending lying | 26.44% | 0 |
| **falling to the left and recovery** | **88.37%** | **42.31%** |
| rolling out of bed and going on the floor | 0 | 0 |
| step off the podium | 0 | 0 |
| falling on the floor vertically | 4.49% | 0 |
| falling down slowly slipping on a wall | 3.41% | 0 |

### D. IMU Placement Positions

Synthetic IMU data was generated from different positions (shank, waist, thigh, chest, head, wrist, ankle). The performance of models trained on these synthetic datasets was compared, and the results were contrasted with models trained on real IMU data. The performance comparison was conducted based on accuracy using the dataset A.

Among the seven regions of interest, the models trained on IMU data from the shank and ankle exhibit the highest accuracy, regardless of whether the data is synthetic or real, with accuracy levels hovering around 90%. However, the models trained on data from other regions generally achieve accuracy rates ranging from 60% to 70%, with the highest not exceeding 80%. It is evident that in Dataset A, the IMU data from the shank region demonstrates notably superior discriminatory power between fall and non-fall actions compared to other positions. Similar testing was conducted on Dataset B, yielding consistent outcomes.

TABLE VI

PERFORMANCE OF LSTM MODELS TRAINED WITH IMU DATA FROM DIFFERENT BODY LOCATIONS ON DATASET A

| IMU placement | simulated | real |
|---|---|---|
| head | 61.84% | 66.83% |
| chest | 70.70% | 76.92% |
| waist | 70.16% | 79.09% |
| thigh | 73.70% | 76.92% |
| **shank** | **91.99%** | **87.50%** |
| **ankle** | **89.39%** | **89.66%** |
| wrist | 65.59% | 63.04% |

### E.  Combination of Multiple IMUs

Combining the synthetic IMU data from multiple body locations within the same experimental trial, an LSTM model was trained and tested on Dataset B. For the scenario involving two IMUs, we considered four candidate body locations (chest, waist, thigh, shank) and explored all possible six combinations.

Among all combinations, the training data combination of waist and shank achieves the highest model performance, reaching 89.09%. In fact, models trained on data combinations that include the lower leg segment all attain high accuracy rates (exceeding 85%). Conversely, the accuracy rates of the three combinations lacking lower leg data fall below 80%. In the case of the combination of lower leg and waist data, its accuracy is enhanced compared to using only lower leg or waist data individually.

TABLE VII

ACCURACY OF MODELS TRAINED WITH COMBINATIONS OF IMUS FROM DIFFERENT BODY LOCATIONS  ON DATASET B

| IMU placement | accuracy |
|---|---|
| shank + tight | 86.12% |
| **shank + waist** | **89.09%** |
| shank + torso | 87.39% |
| thigh + waist | 74.39% |
| thigh + torso | 75.18% |
| waist + torso | 69.35% |

## VI. DISCUSSION

### A. Performance of Models Trained with Synthetic Data

The experimental results demonstrate that LSTM models trained using the synthetic IMU dataset generated by Opensim exhibit good performance on both public datasets, with comparability to models trained using real IMU data of the same sample size. On dataset A, the model trained with synthetic data achieves higher accuracy compared to the model trained with real IMU data, while the opposite trend is observed on dataset B. The differences in performance among models trained with synthetic IMU data indicate the presence of variations, yet the disparities from models trained with real data remain within an acceptable range. Despite the discernible differences between synthetic IMU data and authentic IMU data at the signal level, the utilization of techniques such as human model scaling and randomized IMU pose offsets during the simulation process introduces data augmentation. Consequently, the occurrence of superior generalization performance to unknown subjects in models trained on the synthetic dataset, as compared to models trained solely on data from a single participant (as observed on dataset A), is also logically explicable. It's important to note that real IMU data collection is limited, while synthetic IMU data can be generated indefinitely. Based on this fact, this study further explores the impact of IMU placement positions on fall detection using arbitrarily generated synthetic IMU data. Different combinations of IMU data from various placements were also explored to enhance the accuracy of the model.

### B. IMU Placement for Fall Detection

The results of testing LSTM models trained with both synthetic and real IMU data indicate that the fall detection model trained with data from the shank (lower leg) IMU achieves the highest accuracy. This holds true for both synthetic and real IMU datasets. Considering that the test dataset encompasses a wide range of actions, including common activities and highly confusing actions, and the model used is an effective LSTM model, these results shed lights on that placing the IMU on the shank region provides superior performance for detecting fall actions.

### C. Actions Susceptible to Misclassification

The testing results on dataset B reveal that the model trained using synthetic data correctly classifies the majority of both fall and non-fall actions. However, there are some actions that are particularly prone to misclassification, resulting in high false positive rates.

Among the 16 non-fall actions in dataset B, the actions of transitioning from standing to lying on a bed and from lying to sitting have the highest misclassification rates. These actions involve changes between standing/sitting and lying positions, which have similar movement patterns to fall actions, making misclassification understandable.

Among the 20 fall actions, the actions of falling to the left side and then recovering and falling and quickly recovering have the highest misclassification rates. These actions involve falling and recovering, and their movement patterns are similar to actions where the person stumbles but doesn't fall. Therefore, they are susceptible to being misclassified as non-fall actions. Correctly classifying these highly confusing actions might require the use of alternative models or feature extraction methods. The models trained with real IMU data also show high misclassification rates for these actions, which indirectly validates the reliability of synthetic IMU data.

### D. The Combination of Multiple IMUs

Experimental results on Dataset B indicate that when using two IMUs, the best performance is achieved by combining the IMU data from the shank and waist. This performance also surpasses models trained solely on shank or waist IMU data. This observation suggests that the combination of IMU data from different body locations may enhance the representation of falling actions. The placement of an IMU at the waist captures the motion of the body's center of mass, providing insights into body posture and inclination. Moreover, such positioning usually minimizes user discomfort and interference with natural movements. It is noteworthy that IMU placement at the waist has been commonly adopted in a substantial number of fall detection studies [38]. Conversely, an IMU attached to the lower leg can track motion at the extremity of the body, aiding in detecting localized fall actions or gait anomalies. While utilizing multiple IMUs may enhance fall detection, it's important to consider the inconvenience of wearing numerous IMU devices in daily life. Additionally, this study's model is relatively simplistic and might not fully extract features from multiple IMUs. Exploring methods for the model to better capture the optimal representation of fall actions from multiple IMUs remains an area of investigation.

### E. Quality of Synthetic Data

To assess the quality of synthetic IMU data generated by Opensim from a signal perspective, a comparison was made between synthetic IMU data and real IMU data collected during the same experiments. For accelerometer data, it was observed that real IMU data exhibited a significant amount of noise and spikes, while synthetic IMU accelerometer data appeared smoother. This difference in noise levels could be attributed to the use of Gaussian filtering during the action reconstruction process and the inherent filtering effect of Opensim's inverse kinematics calculation, acting akin to a human motion filter. Furthermore, the discrepancy between synthetic and real IMU data can be partially attributed to the fact that real IMUs can experience shifts on the body's surface due to clothing movement or muscle contractions, whereas simulated IMUs are anchored to specific body parts and do not undergo relative motion. Despite these differences, the synthetic accelerometer data successfully captured the trends and variations of the real IMU data, thus preserving the essential features of falling actions (Fig. 7). Regarding synthetic IMU angular velocity data, it exhibited higher-frequency variations compared to accelerometer data. The quality of generated synthetic IMU data plays a pivotal role in training the model, and there is still room for improvement in the algorithm used to generate high-quality synthetic IMU data. High-quality synthetic data could facilitate the extension of the experimental approach of this study to encompass a broader range of actions, particularly more intense movements like jumping. This expansion further enhances the applicability prospects of the proposed simulation method, such as in the realms of safeguarding children's physical activities [39] and facilitating the rehabilitation of foot drop patients [40].

### F. Using Public Datasets

The principles behind vision-based action reconstruction and motion capture-based action reconstruction are fundamentally the same. Both methods employ least-squares algorithms to optimize the model's motion to best fit the simulated markers' positions to the actual markers' positions. There is a wealth of publicly available motion capture datasets, such as the CMU Motion Capture dataset, that can be used for this purpose. Utilizing these datasets involves performing inverse kinematics calculations.

From publicly available datasets, various common actions such as walking and running can be extracted. If custom actions are required, however, conducting action experiments is still necessary. Moreover, since most existing fall detection research relies on video imagery or wearable IMUs, a substantial number of fall action videos might potentially be employed to reconstruct falling actions. This approach could further reduce the need for extensive real-world fall experiments and contribute to expanding the training dataset with various human motion patterns.

## VI. CONCLUSION

This study investigates the comparability between synthetic IMU data generated by Opensim and real-world data, and demonstrates that the performance of the LSTM model trained on synthetic data aligns with that of models trained on real data. This approach effectively reduces the count of actual fall experiments and mitigates the demands on experimental apparatus, thereby facilitating research in machine learning based fall detection algorithms. This study further investigates the optimization of fall detection based on simulation by exploring the placement of IMUs. The results indicate that the lower leg region is a recommended IMU attachment site, as models trained on data from this position demonstrate the best performance on the test dataset. Finally, the fusion of IMU data from different regions is discussed in terms of its impact on model performance. The findings suggest that when utilizing two IMUs, the placement configuration involving the lower leg and waist regions enhances the ability to distinguish fall actions from non-fall actions. This research provides novel venue for generating low cost synthetic data in scenario where acquiring data for machine learning is challenging and paves the way for customizing machine learning configurations.

## ACKNOWLEDGMENT

The experiments presented in this paper were conducted at the Kaiwu Laboratory of Shanghai Research Institute for Intelligent Autonomous Systems.